\documentclass[usenatbib, useAMS, referee, lscape]{mn2e}

\newcommand{\be}{\begin{equation}}
\newcommand{\bs}{\begin{subequations}}
\newcommand{\cc}{{~\rm cm^{-3}}}
\newcommand{\cm}{{~\rm cm}}
\newcommand{\der}[2]{\frac{d {#1}}{d {#2}}}
\newcommand{\ee}{\end{equation}}

\newcommand{\es}{\end{subequations}}
\newcommand{\g}{{\mathbf g}}
\newcommand{\K}{{~\rm K}}
\newcommand{\keV}{{~\rm keV}}

\newcommand{\kms}{{~\rm km\; s^{-1}}}
\newcommand{\kpc}{{~\rm kpc}}

\newcommand{\psec}{~\rm pc}

\newcommand{\vv}{{\mathbf v}}
\newcommand{\xx}{{\mathbf x}}
\newcommand{\yr}{{~\rm yr}}

\usepackage{amsmath}                   
\usepackage{amssymb}                   
\usepackage[UKenglish]{babel}          
\usepackage[colorlinks=true]{hyperref} 
\usepackage{multirow}                  
\hypersetup{urlcolor=blue}             
\usepackage{graphicx}                  
\usepackage{subfigure}                 
\usepackage{mathrsfs}                  

\title[Angular momentum and cold feedback]
{Solving the angular momentum  problem in the cold feedback mechanism
of cooling flows}

\author[Pizzolato \& Soker]
{
	Fabio Pizzolato
	\thanks{E-mail: fabio.pizzolato@brera.inaf.it}
	\\
	INAF--Osservatorio Astronomico di Brera, Via Brera n.28, 20121 Milano (Italy)
	\and
	Noam Soker
	\thanks{E-mail: soker@physics.technion.ac.il}
	\\ Physics Department, Technion, Haifa 32000 (Israel)
}

\begin{document}
\date{\today}

\pagerange{\pageref{firstpage}--\pageref{lastpage}} \pubyear{2010}

\maketitle

\label{firstpage}


\begin{abstract}
We show that cold clumps in the intra--cluster medium (ICM) 
efficiently lose their angular momentum as they fall in, such that they can 
rapidly feed the central AGN and maintain a heating
feedback process. 
Such cold clumps are predicted by the cold feedback model, a 
model for maintaining the ICM
in cooling flows hot by a feedback process.  The clumps  very effectively  lose
their angular momentum
in two channels: the drag force exerted by the ICM
and the random collisions between clumps  when they are close to the central 
black hole. 
We conclude that the angular momentum cannot prevent the
accretion of the cold clumps,
and the cold feedback mechanism is a viable model for a feedback mechanism in
cooling flows. Cold feedback does not suffer from the severe problems 
of models that are based on the Bondi accretion. 
\end{abstract}


\begin{keywords}
galaxies: clusters: intracluster medium
galaxies: clusters: cooling flows
galaxies: clusters: general
galaxies: clusters: individual: Virgo

\end{keywords}

\section{Introduction}
\label{s-intro}

For more than a decade now it is clear that the intra-cluster medium (ICM)
in cooling flow (CF) clusters of galaxies and CF galaxies must be heated,
and the heating process should be stabilized by a feedback mechanism
(see reviews by \citealp{Pet06} and \citealp{McN07}).
However, in many cases the heating cannot completely offset cooling
(e.g., \citealp{Wis04, McN04, Cla04, Hic05, Bre06, Sal08, Wil09})
and some gas cools to low temperatures and flows inward (e.g., \citealp{Pet06}).
The mass inflow rate is much below the one that would occur without heating, and the flow is
termed a moderate cooling flow 
\citep{Sok01, Sok03, Sok04}.

Two basic modes have been proposed to feed the super massive black hole (SMBH) at
the heart of the active galactic nucleus (AGN) in CF clusters (e.g., \citealp{Com08}).
They can be termed the hot and the cold feedback mechanisms.
In the hot-feedback mode the hot gas from the vicinity of the SMBH is accreted, such
as in the Bondi accretion process (e.g., \citealp{Omm04, All06, Rus10}).
More sophisticated schemes that use direct calculations of the inflow of the hot gas,
but are basically similar to the Bondi accretion process,
also exist (e.g., \citealp{Cio09, Cio10}).
However, the Bondi accretion process suffers from two severe problems
\citep{Sok06, Sok09}, to the point that it fails.
First, there is no time for the feedback to work.
The accretion of gas from the very inner region does not have time to respond to
gas cooling in the outer regions ($r \ga 10 \kpc$).
Second, although in galaxies the accretion rate can in principle account for the AGN power
\citep{All06, Cio09, Cio10}, in CF clusters the Bondi
accretion rate is much too low \citep{Raf06}.
In a recent paper it was argued that the Bondi accretion process fails to maintain feedback
in the process of galaxy formation as well \citep{Sok09b}.

In the cold feedback model  \citep{Piz05, Sok06, Piz07}
the mass accreted by the central black hole originates
in non-linear over-dense blobs of gas residing in an extended region of $r \sim 5 - 30 \kpc$;
these blobs are originally hot, but then cool faster than their environment and sink
toward the centre (see also \citealp{Rev08}).
The mass accretion rate by the central black hole is determined by the cooling
time of the ICM, the entropy profile, and the presence of inhomogeneities 
\citep{Sok06}.
In the cold feedback model if gas cools at large distances, so do the overdense blobs.
In a relatively short time they feed the SMBH, and the ICM is heated on a time scale
shorter than its cooling time scale.
This overcomes the main problem of models based on the Bondi accretion process
\citep{Sok09b}.
\citet{Wil09} suggest that the behaviour and properties of the cold clumps
they observe in the cluster A~1664 support the cold feedback mechanism.
In general, the presence of large quantities of cold ($T \la 10^4 \K$) gas in CF clusters
(e.g., \citealp{Edg02, Edw07}) suggests that the non-linear perturbations
that are required to form the dense clumps do exist in CF clusters.

The main arguments against the cold feedback mechanism was that the cold clumps
supposed to feed the SMBH have too much angular momentum, and they 
cannot fall directly to
the SMBH: for this reason,  their accretion rate will be too slow and the response time
too long (e.g., \citealp{Rus10}).

In this paper we show that dense clumps lose their angular momentum rapidly enough,
such that there is no angular momentum problem in the cold feedback mechanism.
The equations are presented in Sections~\ref{s-eqns} and~\ref{s-angular}, 
and the cluster properties used for
the calculations are given in Section~\ref{s-prof}.
In Section~\ref{s-numerical} we present the solution for falling clumps, 
both without (Sec.~\ref{s-lzero}) and
with (Sec~\ref{s-l}) angular momentum. 
In Section~\ref{s-collision} the loss of angular momentum
via collision is estimated. 
Our main conclusions are in Section~\ref{s-summary}.

\section{Equations of Motion}
\label{s-eqns}

Following \citet{Piz05} (also Paper~I), 
we write the set of equations governing the motion of a
cold clump in the ICM. 
In Paper~I we adopted the following equations for the  motion  of  a clump:
\bs
\begin{align}
\label{e-pos}
&\der{\mathbf x}{t} = \vv
\\
\label{e-newton}
& \rho'\; V \; \frac{d\vv}{dt} = - \frac{C_D}{2} \; S \; \rho \; |\vv|  \; \vv 
+ \g \,(\rho'-\rho) \; V, 
\end{align}
\es
where the primed quantities refer to the clump.
The first equation  relates the position $\xx$ of the clump with its velocity $\vv$;
the second is Newton's second law for the clump's motion: $V$ is the volume of
the clump, and $\rho' \, V$ its mass. The first term on the
right--hand side is the drag force, 
where $C_D\simeq0.75$ \citep{Kai03} is  the drag coefficient and 
$S$ the clump's cross section. The second term
is the gravitational force, corrected for the buoyancy.

\begin{figure}
\begin{center}
\vskip10mm
\includegraphics[width =110mm, angle = 0]{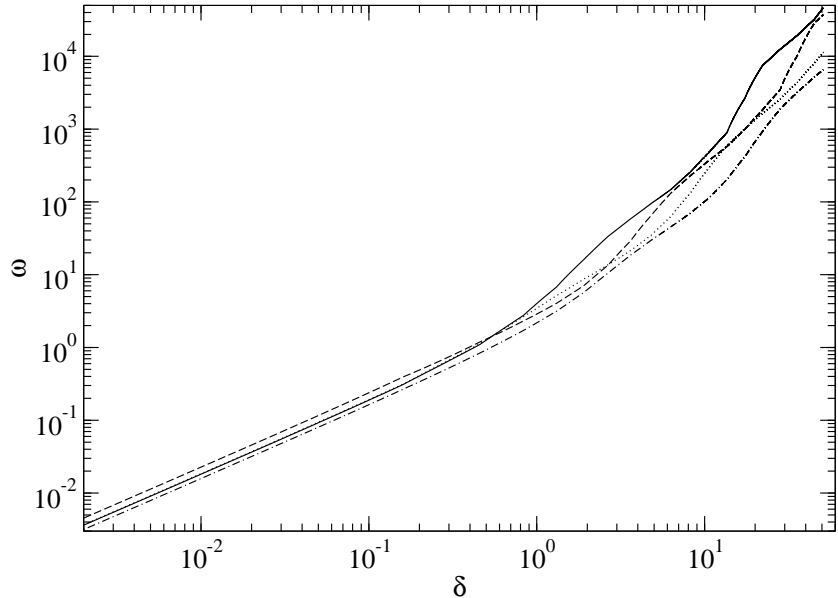} 
\end{center}
\caption{
Plot of the function $\omega(\delta, T)$ defined by Equation~\eqref{e-ft}.
$\delta$ is the density contrast (eq.~\ref{e-dcontrast}),
while $\omega$ is a function that enters the equation describing the relative
evolution of the clump and the ICM;
the larger $\omega$ is, the faster is the evolution of the clumps due to radiative cooling. 
Terms due to buoyancy and to radiative cooling 
appear in $\omega$.
The cooling function used to draw this plot has metal abundance $Z=0.3$ 
(\citealp{And89} solar units).
The curves refer to different temperatures: 
$T=0.5\keV$ (solid line).
$T=1.0\keV$ (dashed line)
$T=2.0\keV$ (dotted line) and
$T=4.0\keV$ (dash--dotted line).
}
\label{f-ft}
\end{figure}

These equations must be completed with the evolution of the density of the clump,
which we give in a slightly different form than in Paper~I.
Let $s(\xx, t)$ and $s'(\xx, t)$ be the specific entropies 
of the ICM and of
a clump at the time $t$ and position $\xx$.
We assume that the clump is always
in pressure equilibrium with the ICM: $P=P'$. 
For this reason the specific entropy $s'$ of an overdense  clump is  
lower than the ICM specific entropy. 
Let
\be
\label{e-deltas0}
\Delta s(\xx, t) = s'(\xx,t) - s(\xx , t)
\ee
be the entropy difference between the clump and the ambient ICM at the time $t$.
We now consider the evolution of $\Delta s$ at a later instant
$t+d t$.
On account of the radiative losses both the ICM and the clump have cooled. 
The entropy profile of the ambient gas  at $t+dt$ is
\be
\label{e-dsa}
s(\xx, t+d t) - s(\xx, t) = - \frac{n_e\, n_H \Lambda(T)}{n\, T}\; dt,
\ee
where $\Lambda(T)$ is the cooling function, $n_e$, $n_H$ and $n$
being the  electron, proton and total density of the ICM
\footnote{
In the present analysis we neglect the ``background'' flow due to the cooling of
the ICM.  The effect of this term was addressed by \citet{Low89}. 
Its inclusion  would require some modelling
of the background flow. The impact on the results is weak, and does  not
justify  the  amount of required work.
}.
At the same instant $t+dt$ the entropy content of the clump has changed as well
due to the radiative losses: 
\be
\label{e-dsc}
s'(\xx +d\xx, t+dt) - s'(\xx, t) = -\frac{n'_e\, n'_H \Lambda(T')}{n'\, T'}\; dt,
\ee
The position is now evaluated at $\xx+d \xx$ because the clump is denser than the 
environment, so it has sunk a little, moving from $\xx$ to 
$\xx + d\xx$. 
The entropy contrast  with the ambient gas at the new position of the clump is
\be
\label{e-deltas1}
\Delta s(\xx+d\xx, t+d t) = s'(\xx+d\xx, t+dt) - s(\xx+d\xx, t+dt).
\ee
Subtracting equations~\eqref{e-deltas1} and~\eqref{e-deltas0}
taking into account equations~\eqref{e-dsa} and~\eqref{e-dsc}
we find  the variation of the entropy contrast:
\be
\label{e-e0}
d \Delta s = \Delta s(\xx+d\xx, t +d t) - \Delta s(\xx, t) =
- \frac{n'_e\, n'_H \Lambda(T')}{n'\, T'}\; dt +
\frac{n_e\, n_H \Lambda(T)}{n\,  T}\; dt - \nabla s\cdot d\xx
\ee
The entropy contrast $\Delta s$ may be expressed in term  of the density
contrast 
\be
\label{e-dcontrast}
\delta = (n' - n)/n
\ee 
under the isobaric assumption $P\equiv P'$
between the clump and the ICM:
\be
\Delta s = \tfrac{3}{2}\, k_{B}\,\ln\left(\frac{P'}{n'^{5/3}}\right)
- \tfrac{3}{2}\, k_{B}\, \ln\left(\frac{P}{n^{5/3}}\right) =
- \tfrac{5}{2}\, k_{B}\, \ln(1+\delta);
\ee
notice that if $\delta=0$ the entropy contrast vanishes.
Equation~\eqref{e-e0} may then be written
\be
\label{e-entropy0}
\tfrac{5}{2} \frac{1}{1+\delta} \der{\delta}{t} =
\tfrac{3}{2}\, \vv\cdot \nabla \ln K  + 
\frac{n'_e\, n'_H \Lambda(T')\; - n_e\, n_H \Lambda(T)}{P},
\ee
where we have introduced  the pressure $P'=P= n k_B T$ and the entropy index 
$K=T/n_{e}^{2/3}$, connected to the specific entropy $s$
through the relation
\be
\label{e-eindex1}
s = \tfrac{3}{2} \: k_{B} \ln K.
\ee
The temperature of the clump is related to that
of the ambient gas through the isobaric condition
\be
\label{e-contrast}
T' = T/(1+\delta).
\ee
With the aid of this constraint we may rewrite Equation~\eqref{e-entropy0} 
\be
\label{e-entropy}
\tfrac{5}{2} \frac{1}{1+\delta} \der{\delta}{t} =
\tfrac{3}{2}\, \vv\cdot \nabla \ln K  +
\frac{\mu_H\;  n_e\; \Lambda(T)}{k_B T} \;
\omega(\delta, T),
\ee
where $\mu_H = n_H/n\simeq 0.44$.
The function
\be
\label{e-ft}
\omega(\delta, T) = 
\frac{(1+\delta)^2}{\Lambda(T)}\; \Lambda\left(\frac{T}{1+\delta}\right) - 1,
\ee
is plotted in Fig.~\ref{f-ft} for several values of $T$. Note that
for small values of $\delta$ the profiles are very similar to each other, since
$\delta\ll1$ implies $\omega \simeq (2-d\ln\Lambda/d\ln T)\,\delta$ for any $T$.

Since relaxed clusters are convectively stable, the entropy gradient 
is positive, and since  $\vv$ is directed inwards, 
the first term on the right--hand side of
Equation~\eqref{e-entropy} is negative, and 
tends to stabilise the clump, i.e. to
keep its density contrast small. 
This term is responsible for the stability of clumps with small density contrasts. 
The second term  describes the clump's instability to its own radiative cooling
with respect to the ICM.  It has the sign of $\omega$, which
is usually positive for $\delta>0$, thus giving a positive contribution to $\dot\delta$.
As discussed in more detail in Appendix~\ref{s-equilibrium}, for small $\delta$
this term is responsible for an overstability of the clump, as already
found and discussed by \citet{Bal89}.

As pointed out by \citet{Bur00}, when the cooling time of the clump
falls below the sound crossing time, the pressure equilibrium does not
hold any longer, and the clump evolves isochorically.
For consistency, we shall have to check {\em a posteriori} that the clump does
not enter this regime during its evolution.

Equation~\eqref{e-entropy} is the same equation for the evolution of
the density contrast $\delta$ adopted in Paper~I if the term with the ICM
cooling is dropped. The new equation expresses more transparently than in
Paper~I the role of the ambient entropy gradient in the clumps' 
accretion \citep{Sok06}.
The dynamical status of the overdensity (the right--hand side
of Equation~\ref{e-entropy}) depends on the properties of the clump
only through its overdensity: the remaining controlling
parameters only depend on the ICM temperature, density and entropy
distributions.
At this stage, we are neglecting the ICM heating due to the AGN activity, and
we only consider its radiative cooling.
The ICM cooling term  makes the clumps slightly more stable than considered in 
Paper~I, where even a clump of zero overdensity gave a positive
contribution  to $\dot\delta$.

\section{The Angular Momentum}
\label{s-angular}

As in Paper~I, we follow the evolution of a clump, but the purpose of this study
is to address more fully the issue of the clump's angular momentum.
If the clump is not able to lose effectively its angular momentum, it
cannot be accreted by the AGN rapidly enough, and the feedback cannot work.

We  estimate the time scale 
taken by the drag force to dissipate the angular momentum of
the clump. Taking the vector product of $\xx$ by Equation~\eqref{e-newton}
we get
\be
\der{\mathbf l}{t} = - \tfrac{3}{8}\;\frac{C_D \; |\vv|}{a \; (1+\delta)} \; \mathbf l
\ee
where $a$ is the radius of the clump (assumed spherical) and 
$\mathbf l = \xx \times \vv$ is the specific angular momentum.  Note that since
$\mathbf l \times \dot{\mathbf l}=0$ the orbit of a clump is planar.
The characteristic time  of angular momentum loss is
\be
\label{e-lloss}
\tau_{l} = {\mathbf l}/{\dot{\mathbf l}} = 
\frac{8}{3} \frac{a\; (1+\delta)}{C_D\; |\vv|}
\simeq
1.2\times 10^7 \yr \;
\left(\frac{a}{100\psec}\right)\;
\left(\frac{1+\delta}{10}\right)\;
\left(\frac{\sigma}{300\kms}\right)^{-1},
\ee
where we have equalled $|\vv|$ to a typical velocity dispersion $\sigma$.
This time is to be compared to the time of a Keplerian revolution  to give
\be
\tau_{l}/\tau_K \simeq (1+\delta) a / r
\ee
For $\delta\simeq 10$, $a\simeq 100\psec$ and $r\simeq 10 \kpc$
this ratio is $\tau_{l}/\tau_K\simeq 10^{-1} \ll 1$. 
Therefore a clump of moderate overdensity
is expected to  quickly lose its angular momentum and fall straight on the AGN.
An exception to this occurs when the cooling time of the clump is much less than 
$\tau_{l}$.
In this case, the clump  becomes very dense ($\delta\gg1$) 
before losing its angular momentum.
For $\delta\gg1$ the drag force is small, and the clump may orbit around the centre
losing but slowly its angular momentum.
This approximate derivation must be complemented  by a  more accurate numerical
calculation, presented in  Section~\ref{s-numerical}.

Since the orbit of a clump is planar, the differential equations for its (isobaric) 
evolution may be written in two--dimensional polar coordinates $(r,\phi)$
\bs
\label{e-evo}
\begin{align}
& \der{r}{t} = u
\\
& \der{\phi}{t} = l/r^2
\\
\label{e-evo-u}
& \der{u}{t} = \frac{l^2}{r^3} - \tfrac{3}{8}\; C_D\; \frac{|\vv|}{a\,(1+\delta)}  
\; u - g \frac{\delta}{1+\delta}
\\
\label{e-evo-l}
& \der{l}{t} = - \tfrac{3}{8} \; C_D\; \frac{|\vv|}{a\,(1+\delta)}  \; l
\\
\label{e-evo-d}
& \der{\delta}{t} =  (1+\delta) \;
\left[\tfrac{3}{5}\, \frac{u}{r} \; \der{\,\ln K}{\,\ln r} + 
\tfrac{2}{5}\,
\frac{\mu_H\, n_{e}\, \Lambda(T)}{k_B T}\;\omega(\delta, T)
\right],
\end{align}
\es
where $u$ is the radial velocity, $l$ the modulus of the angular momentum,
$g =|\mathbf g|$ is the modulus of the gravitational acceleration, and
\be
|\vv| = \left(u^2 + l^2/r^2\right)^{1/2}
\ee
is the modulus of the velocity.

The last equation we need  governs the evolution of the clumps' size $a$.
There are different possible choices here: 
the most straightforward one  implements the 
conservation of the clump's mass
\be
\label{e-mass}
a = a_0 \left[\frac{1+\delta_0}{1+\delta}\, \frac{n_e(r_0)}{n_e(r)}\right]^{1/3},
\ee
where $r_0$ is the initial radial distance of the clump  from the centre, $a_0$
and  $\delta_0$ are the initial clump's radius and overdensity, respectively.
On the other hand, as a clump becomes dense and moves through the ICM, 
it may lose its spherical shape, or even  break up into several smaller clumps.
The fragmentation (or mass loss) is clearly inconsistent with the
conservation law~\eqref{e-mass}, so we shall have to implement the description of 
this phenomenon at some level. The process of fragmentation is quite complex and
beyond the scope of the present paper, so we take  a phenomenological stand, 
and  assume that at each radial distance from the centre $r$  the
mass of the clump  $m$ is reduced with respect to its initial value $m_{0}$
according to the prescription 
\be
\label{e-breakup}
m = m_{0} \; (r/r_{0})^{k},
\ee
where the mass loss index $k\geq0$ is a  constant. 
This prescription is equivalent to 
\be
\label{e-mass2}
a = a_0 \left[\left(\frac{r}{r_{0}}\right)^{k}\frac{1+\delta_0}{1+\delta}\, \frac{n_e(r_0)}{n_e(r)}\right]^{1/3},
\ee
which reduces to the mass conservation~\eqref{e-mass} if  $k=0$.
Note that the decrease of $a$ implied by Equation~\eqref{e-mass2} has the same effect to
increase the cross section to volume ratio, which  mimics the clump's loss of spherical symmetry, with the  major axis aligned with  the direction of motion.

Note that Equation~\eqref{e-breakup} properly describes the effects of 
loss of shape and/or  break up if the clump does not oscillate across the ICM, since in this 
case $m(t)$ would not be a monotonic function of time.

In Section~\ref{s-numerical} we shall solve numerically the 
system of equations~\eqref{e-evo} separately for the prescriptions~\eqref{e-mass} 
and~\eqref{e-mass2}.
In Appendix~\ref{s-equilibrium} we  present an analysis of 
the equilibrium solutions of Equations~\eqref{e-evo} 
(with the conservation condition~\ref{e-mass}),
and also provide an approximate stability criterion for the evolution of
a clump (see also \citealp{Bal89, Tri91}).

\section{Choice of the Cluster Profiles}
\label{s-prof}

\begin{figure}
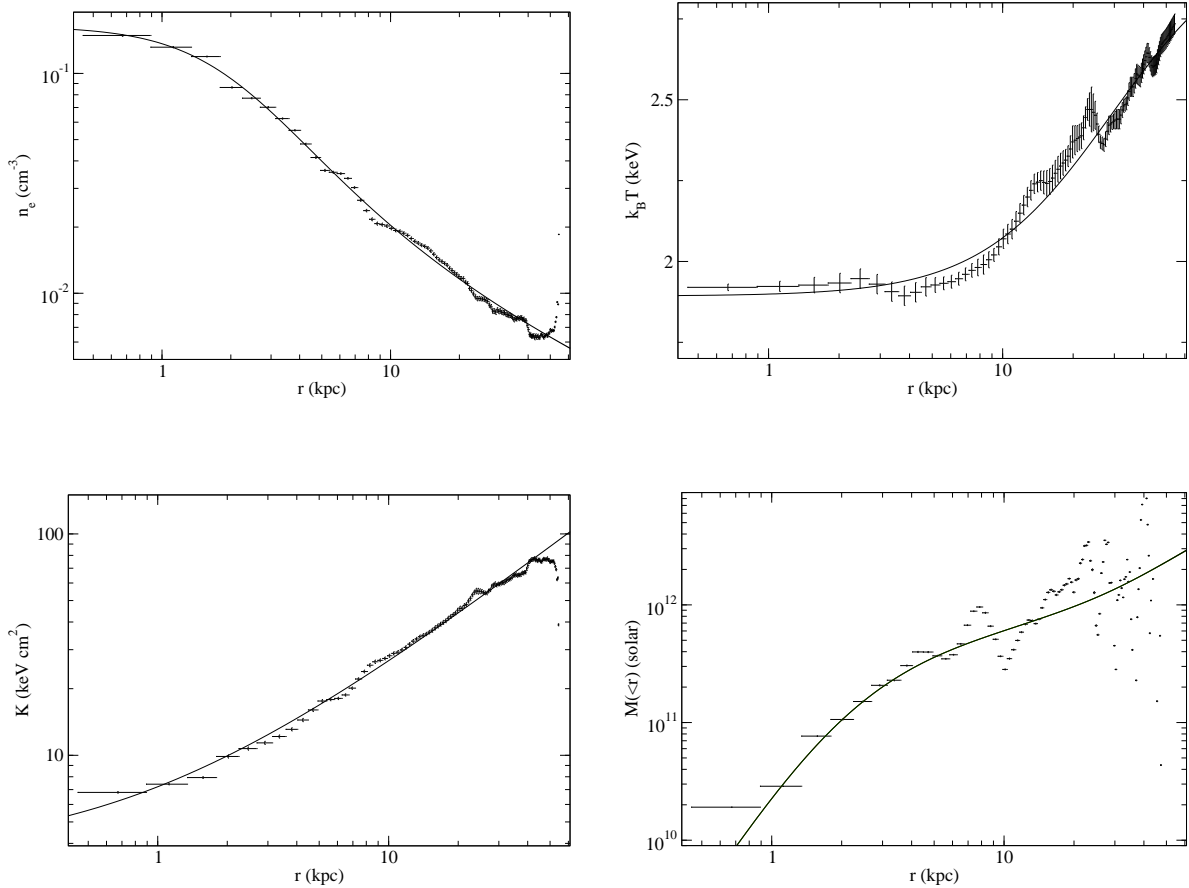

\begin{center}
\vskip10mm
\subfigure{\includegraphics[width =75mm, angle = 0]{fig02a.eps}}
\qquad
\subfigure{\includegraphics[width =75mm, angle = 0]{fig02b.eps}}
\vskip10mm
\subfigure{\includegraphics[width =75mm, angle = 0]{fig02c.eps}}
\qquad
\subfigure{\includegraphics[width =75mm, angle = 0]{fig02d.eps}}
\end{center}
\caption{Best fitting profiles for the electron density (top left), temperature (top right), entropy index (bottom left)  and gravitating mass (bottom right) profiles  adopted for Virgo,
superposed to the data
by~\citet{Cav09}.
}
\label{f-prof}
\end{figure}

As in Paper~I, we focus on a specific cluster, i.e. Virgo. We adopt the
entropy, temperature and electron density profiles  derived by \citet{Cav09}
\footnote{
Their data (for a sample of 239 clusters observed with {\sl Chandra}) 
may be found at\\
\url{http://www.pa.msu.edu/astro/MC2/accept/}.
}.
We fit the 3D electron density  with a  double beta model
\be
\label{e-2beta}
n_{e} = \frac{n_{1}}{[1+(r/r_{1})^{2}]^{\beta_{1}}} + 
\frac{n_{2}}{[1+(r/r_{2})^{2}]^{\beta_{2}}}. 
\ee
The  3D entropy index profiles are modelled by
\citet{Cav09} with the  simple formula
\be
\label{e-eindex2}
K = K_{0} + K_{1} \left(\frac{r}{100 \kpc}\right)^{\alpha}.
\ee
These entropy profiles have been deduced from non--deprojected (2D) temperatures:
according to \citet{Cav09} this  does not affect the determination of the 3D  
entropy profiles appreciably.
The temperature profiles provided by \citet{Cav09} are 2D. 
Since Virgo is almost isothermal (see e.g. \citealp{Ghi04}), we do
not expect that the use of 2D instead of 3D profiles will severely affects our results.
We then use the 2D temperature profile as an estimate of the real 3D profile,
and we fit it with the expression
\be
\label{e-temp}
T = \frac{T_{0} + T_{1} (r/r_{t})^{\gamma}}{1  + (r/r_{t})^{\gamma}},
\ee
already used by \citet{Vik06} to model the  temperature of
galaxy clusters in their  cool cores.
The profiles~\eqref{e-2beta} and~\eqref{e-temp} are used to determine the
hydrostatic gravitational acceleration 
\be
g = \frac{G\, M(<r)}{r^{2}} = \frac{k_{B}\,T}{\mu\:m_{u}\, r} \; \left(
\der{\,\ln n_{e}}{\,\ln r} + \der{\,\ln T}{\,\ln r}\right),
\ee
where $M(<r)$ is the gravitating mass within the radius $r$, 
$m_{u}$ is the atomic mass unit and $\mu\simeq 0.6$ is the mean  molecular weight.
Table~\ref{t-values} presents the best--fitting values of the fit parameters
for the electron density, temperature and entropy profiles. 
The profiles for Virgo of $T$, $n_{e}$, $K$
and $M(<r)$ are plotted in Fig.~\ref{f-prof}, compared to the original data
from \citet{Cav09}.

Throughout this paper we assume that the metal abundance is
$0.3$ times the solar value \citep{Cav09}, 
with the heavy element ratios taken from \citet{And89}.

\begin{table}
\begin{minipage}{160mm}
\begin{tabular*}{0.75\textwidth}{@{\extracolsep{\fill}} | c | c | c | c | c | c | }
\hline
\multicolumn{6}{| c |}{2D temperature}    \\
\hline
$T_{0}$      &  $T_{1}$      &  $r_{t}$     & $\gamma$    &  &    \\   
($\keV$)     &  ($\keV$)     &  ($\kpc$)    &             &  &   \\
\hline
1.893        &  3.022        &  29.36      &  1.543     &  &   \\
\hline\hline
\multicolumn{6}{| c |}{3D electron density}  \\
\hline
$n_{1}$        & $r_{1}$    & $\beta_{1}$  & $n_{2}$        & $r_{2}$    & $\beta_{2}$  \\
($10^{-2}\cc$) & ($\kpc$)   &              & ($10^{-2}\cc$) & ($\kpc$)   &              \\
\hline
1.320          & 8.762      & 0.24          & 15.08          &  1.896     &  0.819     \\
\hline\hline
\multicolumn{6}{| c |}{3D entropy index}\\
\hline
$K_{0}$           &  $K_{1}$          & $\alpha$    &     &  &  \\
($\keV \cm^{2}$)  &  ($\keV \cm^{2}$) &             &     &  &   \\
\hline
3.53            &  146.6            &  0.8        &   &   &   \\
\hline\hline
\end{tabular*}
\caption{
\label{t-values}
Best fitting parameters for the (projected) temperature, and 3D electronic density and entropy index for Virgo.  More details in the text.
}
\end{minipage}
\end{table}

\section{Numerical Calculations}
\label{s-numerical}

In this section we  solve numerically equations~\eqref{e-evo} with the temperature, entropy and density profiles 
presented in Section~\ref{s-prof}. 
We adopt the embedded Runge-Kutta Prince-Dormand (8,9) method
(e.g., \citealp{Pre07}) implemented in the GNU scientific library
\footnote{
\url{http://www.gnu.org/software/gsl/manual/html\_node/Ordinary-Differential-Equations.html}.
}.
The cooling function $\Lambda$ has been calculated with the {\sc XSPEC}
package for the {\sc APEC} atomic  tables.

In the next two subsections we present the evolution of the clumps 
with and without  angular momentum.

\subsection{Zero Angular Momentum}
\label{s-lzero}

We follow the evolution of a clump of initial radius $a_{0}=100\psec$ 
released from rest at a distance $r_0=20\kpc$ from the cluster's centre.
We distinguish the case where the clump conserves its
mass during the inflow, from the case where the clump undergoes fragmentation.

\subsubsection{Constant Mass Clumps}
\label{s-constant}

The size of a constant mass clump is given, instant by instant, by the
conservation condition~\eqref{e-mass}.
We may distinguish three regimes for the accretion of such a clump:
stable, relatively unstable and absolutely unstable.

\begin{figure}
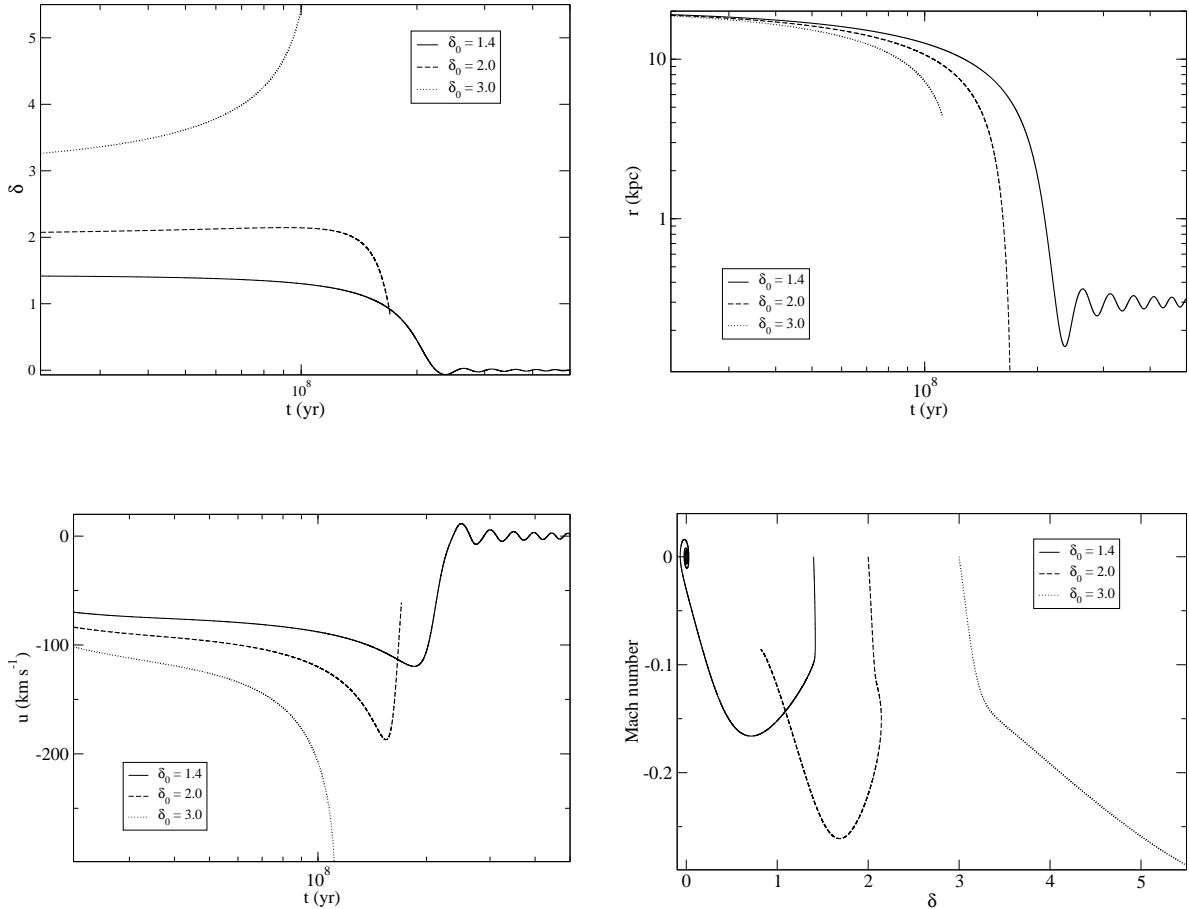

\begin{center}
\vskip10mm
\subfigure{\includegraphics[width = 75mm, angle = 0]{fig03a.eps}}
\qquad
\subfigure{\includegraphics[width = 75mm, angle = 0]{fig03b.eps}}
\vskip10mm
\subfigure{\includegraphics[width =75mm, angle = 0]{fig03c.eps}}
\qquad
\subfigure{\includegraphics[width =75mm, angle = 0]{fig03d.eps}}
\end{center}
\caption{Tracks of a clump with an initial radius $a=100\psec$ released from
$r_{0}=20\kpc$ from the centre.
The lines refer to different initial overdensities:
$\delta_{0}=1.4$ (solid line), 
$\delta_{0}=2.0$ (dashed line) and
$\delta_{0}=3.0$ (dotted line).
Top left: overdensity vs. time;
top right: radial distance vs. time;
bottom left: fall velocity vs. time;
bottom right: Mach  number of the radial fall velocity vs. overdensity.
}
\label{f-tracks}
\end{figure}

These regimes  are exemplified in  Fig.~\ref{f-tracks},
showing  the evolution of  a clump for  different
initial values of the overdensity: 
$\delta_{0}=1.4$ (solid line), 
$\delta_{0}=2.0$ (dashed line) and
$\delta_{0}=3.0$ (dotted line).
The top left panel of Fig.~\ref{f-tracks} shows the evolution of
the overdensity defined by Equation~\eqref{e-dcontrast}. 
The clump with $\delta_{0}=1.4$ is stable. Fig.~\ref{f-tracks} shows that
its overdensity soon drops to zero, oscillating around zero 
with the Brunt--V\"ais\"al\"a
frequency. The clump eventually reaches  an equilibrium position at
$r\simeq0.3\kpc$. The trajectory of this clump in the
($\delta-\mathcal M$) plane (where $\mathcal M$ is the ratio of the radial
velocity to the local sound speed, i.e. the Mach number)
winds up around the origin, which is eventually reached when the energy of the oscillations is dissipated by the drag force.
The case of $\delta_{0}=2.0$ is an example of a  relative
instability: the overdensity decreases,  but  the 
clump is eventually accreted, since it reaches 
the centre of the cluster before  being thermally stabilised.
If the overdensity exceeds  a critical value  $\delta_{C}\simeq2.4$
(slightly less than the estimate $\delta_{C}=1.7$ 
calculated in Appendix~\ref{s-equilibrium}) 
the overdensity always increases, and the clump is eventually accreted 
to the centre:  this is the absolute instability, shown 
in Fig.~\ref{f-tracks} by the case $\delta_{0}=3.0$.

The isobaric prescription for the evolution of very dense 
(i.e., absolutely unstable) clumps
breaks down when  the sound crossing time of the clumps becomes longer than 
its cooling time $\tau^\prime_{\rm sound} > \tau^\prime_{\rm cool}$ 
\citep{Bur00}.  
For our  clump with  $\delta_{0}=3.0$ this happens  when 
$\delta\simeq 32$, or $T'\simeq 0.06\keV$.
The growth of the overdensity stalls after the equivalence 
$\tau'_{\rm cool} \sim \tau'_{\rm sound}$ has been reached,  
and from this time on the clump evolves isochorically.
The subsequent evolution of the clump's temperature 
is difficult to estimate, but we may note that the cooling function
sharply drops below $T\sim 10^{4}\K$ (see e.g. \citealp{Spi78, Sut93}), and 
since $\tau'_{\rm cool}$ greatly increases, the clump
is unlikely to cool much below this limit before being eventually accreted.
The conclusion is that the temperature of the clumps feeding the AGN may be
$T'\sim 10^{4}\K$,  but probably not much colder.

On the other hand, a very dense clump  may 
undergo a gravitational instability: it   breaks up, 
decouples from the ICM and eventually 
freely falls to the centre. 
This evolution is rapid, and it may be possible that the 
gas of the clump  encounters 
an atomic or  molecular phase, as suggested by  the
presence of a large amount of molecular clouds far from the centre of
clusters \citep{Sal08}. In this case,  it is
likely that the gas will cool much below  $1000\K$,
possibly down to few tens $\K$.

We have neglected the feedback activity of  the AGN, that heats the ICM, but not
the clumps. This activity increases the ICM entropy, as well as the entropy contrast
between the clump and the ICM. This effect makes the clumps more unstable than found here.

The results obtained thus far are in good agreement with those presented in
Paper~I. The small differences are due to our different choices of
the metal abundance and the cluster's temperature, density and
entropy profiles.

\subsubsection{Effects of Fragmentation}
\label{s-frag}

\begin{figure}
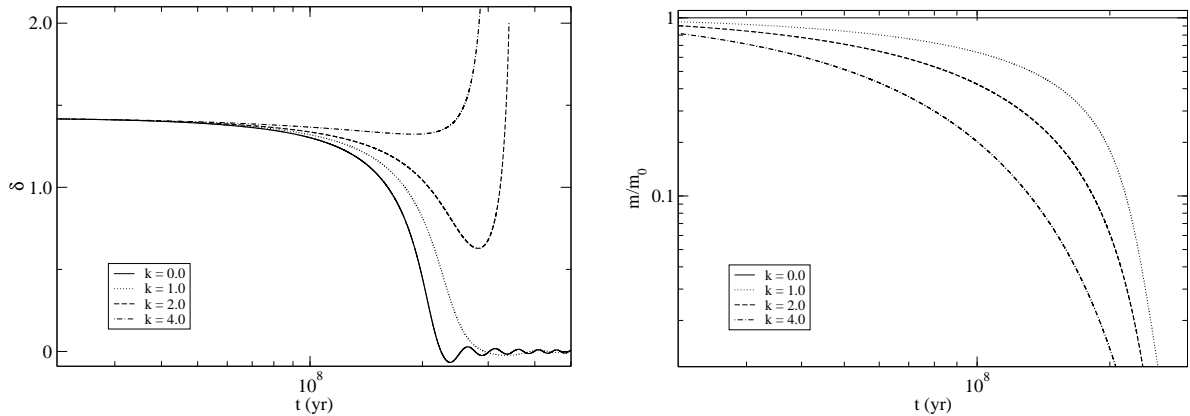

\begin{center}
\vskip10mm
\includegraphics[width =75mm, angle = 0]{fig04a.eps}
\qquad
\includegraphics[width =75mm, angle = 0]{fig04b.eps}
\end{center}
\caption{
Evolution of the overdensity (left panel)
and the mass $m$ of a clump with respect to its initial value
$m_{0}$ (right panel) 
for different values of
the mass loss index $k$ (eq.~\ref{e-breakup}): 
a constant mass clump 
($k=0.0$; solid line),
$k=1.0$ (dotted line)
$k=2.0$ (dashed line) and
$k=4.0$ (large fragmentation of clump; dash-dotted line).
The clump has the initial density contrast 
$\delta_{0}=1.4$,
radius $a_{0}=100\psec$ and is released with zero
velocity and angular momentum from $r_{0}=20\kpc$.
}
\label{f-kdelta}
\end{figure}

We include the description of the clumps' 
break up by setting the mass loss index $k$ 
(Equations~\ref{e-breakup} and \ref{e-mass2})
to a positive value. Fig.~\ref{f-kdelta} compares 
the evolution of $\delta$ for clumps with different $k$,
but the same initial radius ($a_{0}=100\psec$)  and overdensity
($\delta_{0}=1.4$); all the clumps are released from 
rest at the distance $r_{0}=20\kpc$ from the centre.
The curves in the plot refer to the mass loss indexes 
for  the values $k=0$, $k=1.0$, $k=2.0$ and $k=4.0$.
The most relevant effect of a non--vanishing $k$ is a strong
increase of the relative drag force, which  suppresses the oscillations for
$k\gtrsim 2$. 

\begin{figure}
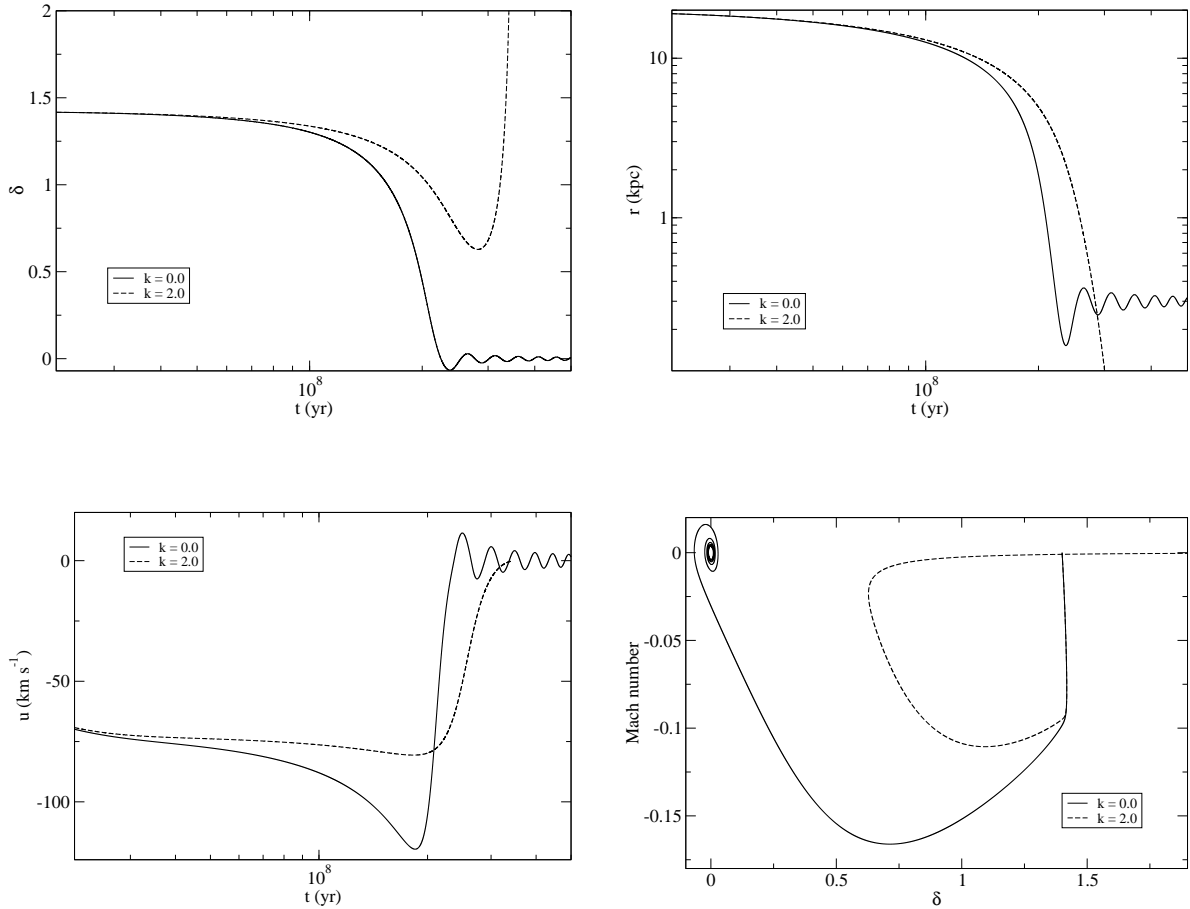

\begin{center}
\vskip10mm
\subfigure{\includegraphics[width =75mm, angle = 0]{fig05a.eps}}
\qquad
\subfigure{\includegraphics[width =75mm, angle = 0]{fig05b.eps}}
\vskip10mm
\subfigure{\includegraphics[width =75mm, angle = 0]{fig05c.eps}}
\qquad
\subfigure{\includegraphics[width =75mm, angle = 0]{fig05d.eps}}
\end{center}
\caption{Dashed line: tracks of a clump with mass loss index $k=2.0$.
The  initial radius and overdensity $a_{0}=100\psec$, 
$\delta_{0}=1.4$. The clump is  released from
$r_{0}=20\kpc$ with zero initial  velocity and
angular momentum.
Top left: overdensity vs. time;
top right: radial distance vs. time;
bottom left: fall velocity vs. time;
bottom right:  fall Mach  number vs. overdensity.
The solid lines show the evolution of a clump with the same
initial  parameters,
but with $k=0$. The case $k>0$ favours the accretion process. 
}
\label{f-ktracks}
\end{figure}

\begin{figure}
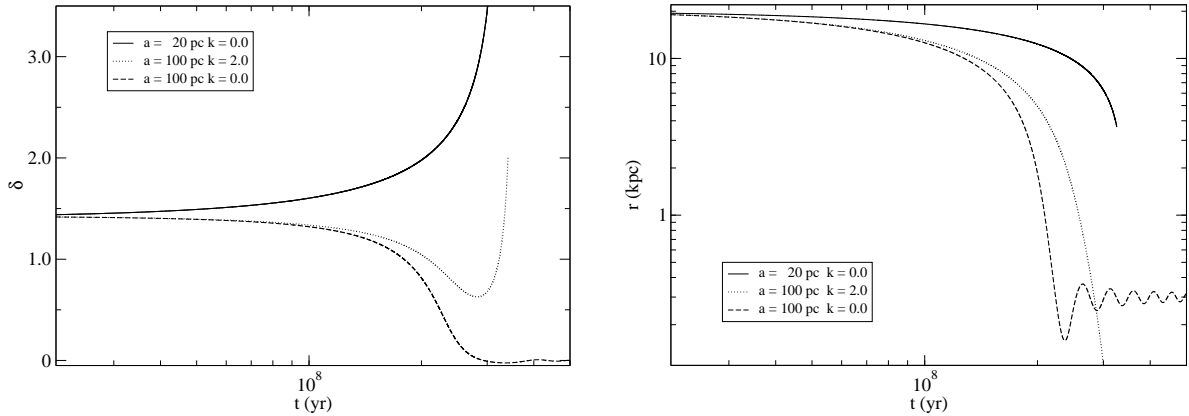

\begin{center}
\vskip10mm
\subfigure{\includegraphics[width =75mm, angle = 0]{fig06a.eps}}
\qquad
\subfigure{\includegraphics[width =75mm, angle = 0]{fig06b.eps}}
\end{center}
\caption{
Evolution of overdensity (left panel) 
and distance from the centre (right panel)  of  clumps 
of different sizes and mass loss indices.
The solid and the dashed line show the evolution of
clumps with $a_{0}=20\psec$ and 
$a_{0}=100\psec$, respectively. For both these clumps it is $k=0$.
For comparison, the dotted line shows the evolution of a clump with 
$k=2.0$ and
$a_{0}=100\psec$.
In all cases the initial overdensity is $\delta_{0}=1.4$, 
and the clump is  released from
$r_{0}=20\kpc$ with zero initial  velocity and
angular momentum.
Smaller clumps fall slower, and therefore their density contrast has time to increase. They are less stable and more prone to  the accretion process.
}
\label{f-atracks}
\end{figure}

Fig.~\ref{f-ktracks} presents the evolution of a clump with 
$\delta_{0}=1.4$ and mass loss index $k=2.0$. For comparison, we have also 
drawn the tracks of a clump with the same parameters, but $k=0$.
If $k=2.0$ the drag force suppresses  the 
oscillations and limits the fall velocity. The clump falls monotonically to 
the centre, and the initial decrease of  its overdensity is reversed. 
We do not present the plots for denser clumps 
since the general pattern remains the same:
the stronger drag force  accelerates the evolution of the clump somewhat,
with  no other relevant qualitative effects.

Fig.~\ref{f-atracks} shows the effect of the size $a$ of the clump  on its evolution.
Since the drag force
behaves as $F_D \propto a^{-1}$, it effectively damps the  oscillations of  
small clumps, whose distance from the centre then decreases monotonically. 
The ratio between the drag force and the force of gravity 
(for the same density contrast) is larger for smaller
clumps. Therefore, the trend of decreasing clump's radius is like that of breaking
($k>0$). 
Smaller clumps fall slower, and therefore their density contrast has time to
grow, making these clumps less stable than the larger ones. 
On the other hand, we also expect that thermal conduction may play  a role
in the evolution of small clumps.
As discussed in Paper~I, we expect that the clumps are protected against the 
evaporation by magnetic fields, which suppress the thermal conduction 
with the ICM by a factor~$\sim10^{-3}$ \citep{Nip04} with respect to its nominal
value (as given by \citealp{Bra65} or \citealp{Spi62}).  Magnetic shielding, however,
becomes inefficient for very small clumps, that will then be evaporated.

According to our phenomenological recipe~\eqref{e-breakup}, 
for moderate values of $k$
the clumps may reach the distance  $r\simeq1\kpc$ still retaining an appreciable
fraction of their initial mass.
Although a more refined analysis is required to make reliable quantitative predictions,
we are confident that this result is valid in general. 
At smaller  radii (below $r\simeq1\kpc$)
the evolution of the clumps is different from the scenario sketched so far,
as discussed  later in Section~\ref{s-collision}.

\subsection{Clumps with Angular Momentum}
\label{s-l}

\begin{figure}
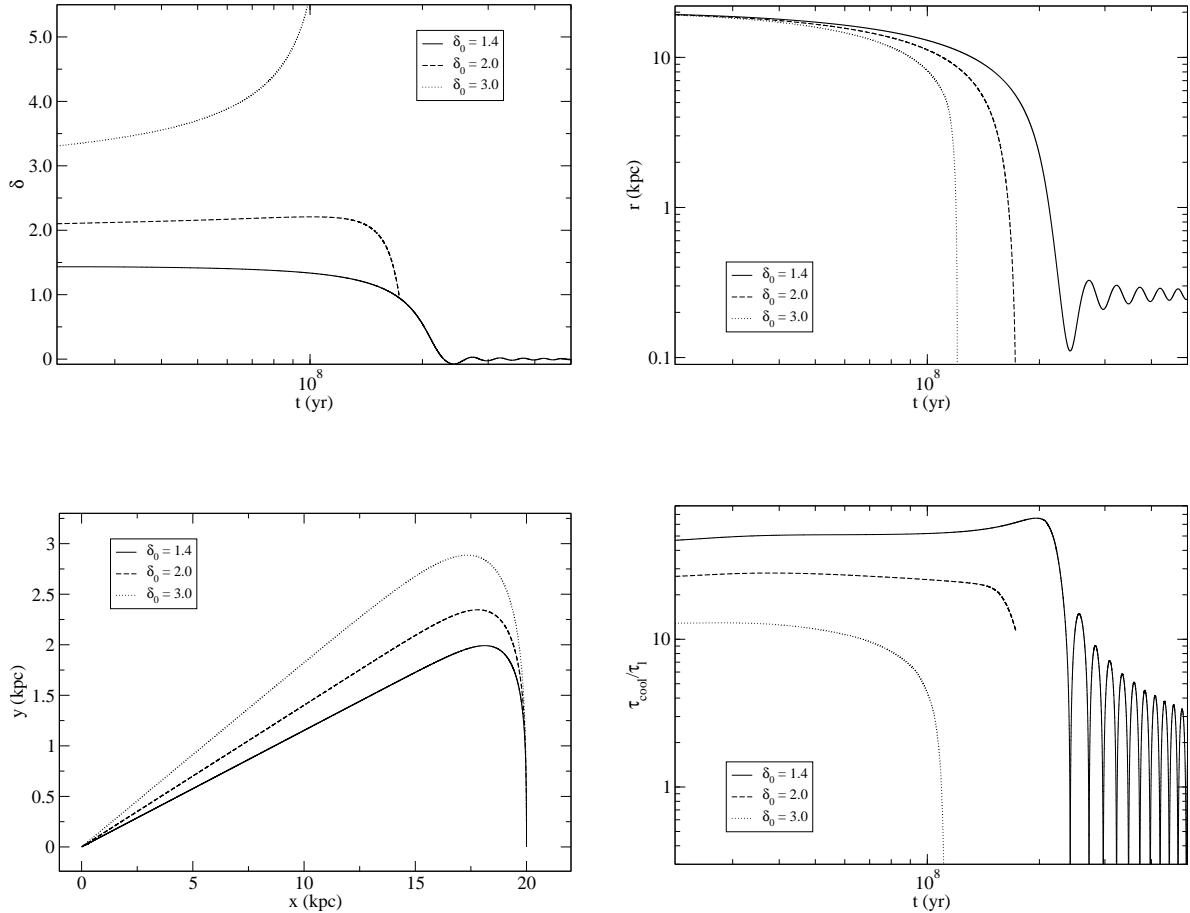

\begin{center}
\vskip10mm
\subfigure{\includegraphics[width =75mm, angle = 0]{fig07a.eps}}
\qquad
\subfigure{\includegraphics[width =75mm, angle = 0]{fig07b.eps}}
\vskip10mm
\subfigure{\includegraphics[width =75mm, angle = 0]{fig07c.eps}}
\qquad
\subfigure{\includegraphics[width =75mm, angle = 0]{fig07d.eps}}
\end{center}
\caption{Tracks of a clump with initial radius $a_{0}=100\psec$ released from
$r_{0}=20\kpc$ from the centre and having an initial 
angular momentum according to Equation~\eqref{e-lcirc}.
The lines refer to different initial overdensities:
$\delta_{0}=1.4$ (solid line), 
$\delta_{0}=2.0$ (dashed line) and
$\delta_{0}=3.0$ (dotted line).
Top left: overdensity vs. time;
top right: radial distance vs. time;
bottom left: shape of the  clumps' trajectories in the orbital plane
(the cluster  centre is at $(x, y) = (0,0)$);
bottom right:  evolution of the ratio between the clump's 
cooling time and the angular momentum loss time.
}
\label{f-jtracks}
\end{figure}

\begin{figure}
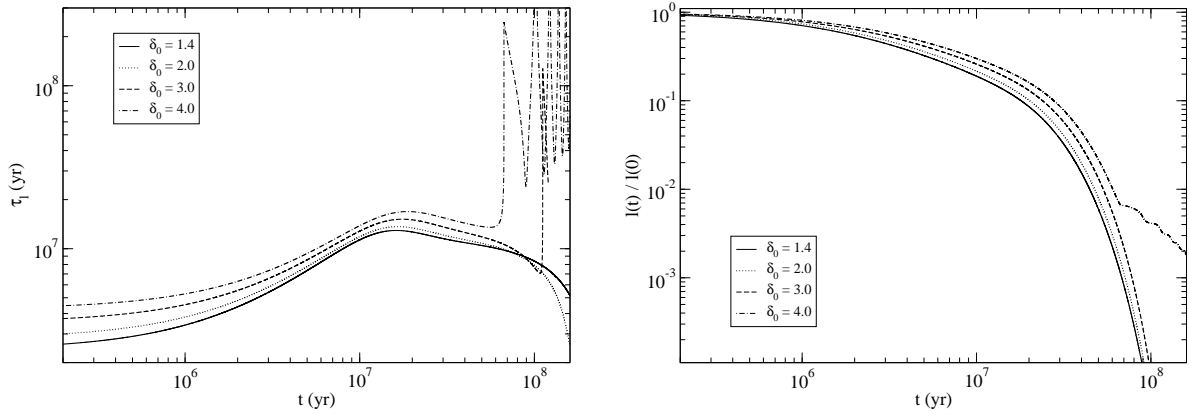

\begin{center}
\vskip10mm
\includegraphics[width =75mm, angle = 0]{fig08a.eps}
\qquad
\includegraphics[width =75mm, angle = 0]{fig08b.eps}
\end{center}
\caption{Evolution of the angular momentum of
a clump with initial radius $a_{0}=100\psec$ released from
$r_{0}=20\kpc$ from the centre and having initial 
angular momentum as in  Equation~\eqref{e-lcirc}.
Left panel: the evolution of the angular momentum loss time
(given by Equation~\ref{e-lloss}); right panel: 
ratio of the angular momentum $l(t)$ with respect to its initial
value $l(0)$.
The lines refer to different initial overdensities:
$\delta_{0}=1.4$ (solid line), 
$\delta_{0}=2.0$ (dotted line).
$\delta_{0}=3.0$ (dashed line) and
$\delta_{0}=4.0$ (dot--dashed line).
The spikes in the curve $\delta_{0}=4$ are due to the
modulation of the clump's velocity on an eccentric orbit.
}
\label{f-jtimes}
\end{figure}

We now consider the accretion  of clumps endowed with an initial orbital
angular momentum. We solve the full set of equations~\eqref{e-evo},
and set the initial angular momentum to 
\be
\label{e-lcirc}
l_{0} = \left[ g_{0}\;  r_{0}^{3}\; \frac{\delta_{0}}{1+\delta_{0}} \right]^{1/2},
\ee
for which  a clump with zero initial radial velocity also has zero radial 
acceleration (see equation~\ref{e-evo-u}, 
where $g_{0}=g(r_{0})$).
The clumps have the same parameters as in  Section~\ref{s-lzero}, i.e.
$a_{0}=100\psec$, $r_{0}=20\kpc$ and  overdensities
$\delta_{0}=1.4$, $2.0$ and $3.0$. A summary of the evolution of
some relevant parameters is plotted in Fig.~\ref{f-jtracks}.
The evolution of the overdensity is not significantly different from 
that found with zero angular momentum for all the clumps. 
As seen from the shape of the orbit 
(bottom left panel of Fig.~\ref{f-jtracks})
and from  the angular momentum loss time (Fig.~\ref{f-jtimes})
all the clumps 
quickly lose their angular momentum, and their  orbit becomes a straight line
pointing to the centre.

\begin{figure}
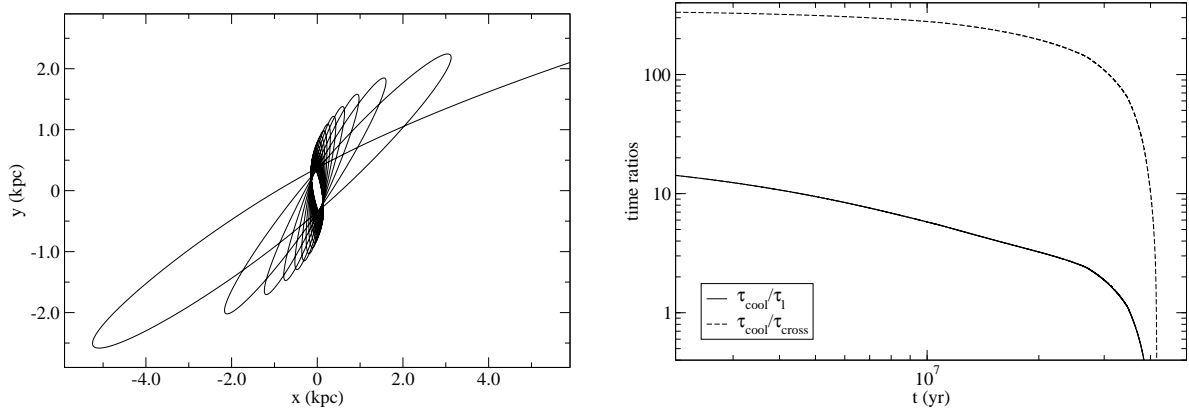

\begin{center}
\vskip10mm
\subfigure{\includegraphics[width =75mm, angle = 0]{fig09a.eps}}
\qquad
\subfigure{\includegraphics[width =75mm, angle = 0]{fig09b.eps}}
\end{center}
\caption{Evolution  of a dense ($\delta_{0}=5$)
clump with initial radius $a_{0}=100\psec$ released from
$r_{0}=20\kpc$ from the centre and having  initial 
angular momentum as in Equation~\eqref{e-lcirc}.
Left panel: the trajectory in the orbital plane.
Right panel: 
ratio  of the cooling and angular momentum loss time (solid line);
ratio of the cooling and sound crossing time (dashed line).
}
\label{f-jdense}
\end{figure}

The clump with $\delta_{0}=1.4$ is still stable:
it reaches an equilibrium radius $r_{\rm eq}\simeq0.3\kpc$ and oscillates around it. 
The denser clumps quickly become very overdense and are eventually accreted 
by the central black hole. 
The last panel of   Figure~\ref{f-jtracks} shows that the clumps 
loose their angular momentum {\em before} cooling.
For even denser clumps this may not be the case. Fig.~\ref{f-jdense}
presents the case of a clump with initial overdensity $\delta_{0}=5$.
After an initial phase during  
which also this clumps loses angular momentum, 
the orbit starts to wrap around the centre
$(x,y)=(0,0)$, with decreasing average distance. 
This clump, unlike the less dense clumps, 
cools down before losing its angular momentum, 
so its overdensity $\delta$ soon grows
very large. The drag force (proportional to $(1+\delta)^{-2/3}$)
becomes very small, increasing $\tau_{l}$ and delaying
the eventual accretion to the centre.  
Albeit  formally correct, this solution may be unphysical for the
following reasons.
First, the orbital  periastron of such dense clumps is 
small (below $100\psec$). As explained in Section~\ref{s-collision},
at these radii the collisions with other clumps  cause the
clump to efficiently lose its orbital angular momentum.
This accelerates the accretion 
process as required by the cold feedback mechanism. 
Second, when the overdensity value of $\sim 45$ is reached 
(for this example, this occurs at 
$t\simeq 4.1\times 10^7 \yr$,  when the clump is still
$r\simeq 16\kpc$ away from the centre), 
the cooling becomes  isochoric rather than isobaric \citep{Bur00}. 
The clump's compression lags behind cooling  and leaves the size
of the clumps larger, increasing somewhat the drag force
relative to what our calculations assume at this phase.

We may summarise the main results of this section as follows: 
the angular momentum is quickly lost by many clumps, and their behaviour is
not much different from the case with zero angular momentum presented in Paper~I
and in Section~\ref{s-lzero}.
This result removes the main obstacle that stood in the way
of the cold feedback mechanism. Namely, we have shown that the response of the AGN
accretion to cooling at large distances is as rapid as required.

\section{Collisional Braking}
\label{s-collision}

In this section we estimate the role of collisions in the innermost region of the cluster.
As the clumps fall in, they are likely to break up into smaller clumps and/or lose their
spherical shape.
Clumps that break up into very small pieces  are likely  to be evaporated by thermal  conduction \citep{Nip04}.
On the other hand, clumps larger than $\sim 10-100 \psec$ 
(for a thermal conduction suppression factor $10^{-2}-10^{-3}$,  Fig.~3 of 
\citealp{Nip04})
will survive, and their subsequent evolution is likely dominated by mutual collisions, as explained in this section.
Loss of spherical shape and fragmentation 
increase the cross section to volume ratio of each clump.
With the higher clumps' number density near the centre, collisions between clumps are
bound to occur.

Consider the clumps  within a sphere $S$ of radius $r$ centred on the AGN. 
The collision time between clumps here  is
\be
\tau_{\rm coll}^{-1} = \sigma \; v_{\rm rel} \; n_c
\ee
where $v_{\rm rel}$ is the relative velocity of two clumps, $n_c$ the number of
clumps per unit volume within $S$, 
and $\sigma$ is the collisional cross section. 
We take it  to be the geometric cross section: $\sigma = \pi (2\, a)^2$,
where the extra factor $2$ comes from the finite size of both the target
and the bullet clump.
This time is to be compared with the clump's crossing time of $S$
\be
\tau_{\rm cross} \sim r / v_{c}, 
\ee
where $v_{c}$ is a typical clump's  velocity.
Random collisions are frequent if $\tau_{\rm coll}<\tau_{\rm cross}$, 
and may efficiently cause the clumps within $S$ to lose their angular
momentum. The ratio
$\tau_{\rm coll}/\tau_{\rm cross}$ therefore estimates the efficiency of the
collisional loss of angular momentum; it may be expressed in terms of the clumps'
volume filling factor  $\epsilon_{V}=4\,\pi a^{3} n_{c}/3$ as
\be
\frac {\tau_{\rm coll}}{\tau_{\rm cross}} \simeq 
\frac{v_{c}}{v_{\rm rel}}  
\frac{a}{r} \frac{1}{3\,\epsilon_{V}}
\simeq
\frac{a/r}{3\, \epsilon_{V}}, 
\ee
where we have assumed  $v_{c}\simeq v_{\rm rel}$ in the last equality.
For example, if $\sim10\%$ of the clumps lose their angular momentum by 
collisions,  then this ratio should be  $\sim10$. 
For $r/a=300$ (say, $r=30\kpc$ and $a=100\psec$, or 
$r=3\kpc$ and $a=10\psec$) the required efficiency 
demands   $\epsilon_{V}\sim 10^{-4}$.
Namely, it is sufficient that only $0.01\%$ of the inner volume is occupied by dense 
clumps in the inner region. In practice, we expect $\epsilon_{V}$ to be larger, e.g.
$10^{-3}-10^{-2}$,  depending on the mass accretion rate.

We conclude that even a small filling factor is enough to make the collisional loss
of angular momentum an efficient process. 
The overall picture is that the drag force dominates the angular momentum loss at large
radii ($r\gtrsim 1\kpc$). At smaller radii ($r\lesssim 1\kpc$)
collisions dominate, causing the clumps  to lose their residual orbital angular momentum.

\section{Conclusions and Summary}
\label{s-summary}

We have studied the role of the cold clumps' angular momentum in
the framework of the cold feedback model, generalizing the results of 
Paper~I that did not include angular momentum.
We first studied falling clumps with zero angular momentum.
The efficiency of accretion increases (i.e., clumps are less stable) as the 
size of the clump
decreases (for a given density contrast) and its cross section increases
(for a given volume and density contrast), as evident from 
Figs.~\ref{f-kdelta}-\ref{f-atracks}.
The reason is that a relatively large cross section implies slower infall, such that
the clump has time to increase its overdensity due to radiative cooling.

We then include angular momentum.
As it is apparent from Figs.~\ref{f-jtracks} and~\ref{f-jtimes}, 
at large radii the drag force is the main agent
cause for clumps to lose their angular momentum, and it is efficient
enough to remove the centrifugal barrier of accreted clumps.
The clumps can easily reach distances of $r \la 100 \psec$.
At this stage, the clumps may be very dense and cool ($T \la 10^4 \K$), and they are
no more in pressure equilibrium with the surrounding ICM.
As discussed in Section~\ref{s-collision}, the collisions between clumps become important,
and even for small volume filling factors ($\epsilon_{V} \simeq 10^{-4}-10^{-2}$) the
collision time scale is comparable or shorter than the infall time.
The residual orbital angular momentum is lost, and the clumps are free
to accrete on the central galaxy and ultimately 
feed the SMBH sitting at at the clusters' centre.
This accretion powers the AGN and activate an efficient feedback with
the ICM cooling.

In our calculations the ICM was static.
However, along the jets axis there is an outflow,
and clumps will not fall there freely. We expect most of the clumps to be accreted from
the equatorial plane. The interaction of the jets and rising bubbles with dense clumps will
be the subject of a future study.
We also note that although the AGN is the main heating source, the falling clumps
do release gravitational energy \citep{Fab03}.

The solution of the angular momentum problem has removed the main theoretical
objection to the cold feedback mechanism as a viable model for ICM heating
in CF clusters. Observational support to the cold feedback mechanism
(e.g., \citealp{Wil09}) stand now on a more solid ground.

\section*{Acknowledgments}

We are grateful to Philippe Salom{\'e}
and Fran\c{c}ois Combes for their valuable comments.
This research was supported by the Asher Fund for Space
Research at the Technion and a grant from the  Israel Science Foundation.


\clearpage
\nocite{}
\bibliographystyle{mn2e}
\bibliography{refs}


\clearpage

\appendix

\section{Equilibrium Solutions}
\label{s-equilibrium}

In this appendix we the study the equilibrium solutions of
equations~\eqref{e-evo}
for a clump with constant mass.
At equilibrium the right--hand sides of~\eqref{e-evo} vanish, which happens if $v=0$, 
$l=0$, $\delta=0$;  $r$ and  $\phi$ are both constant. 
It is clear that since $\delta=0$ the clump is indistinguishable
from the ICM, so this equilibrium solution describes a
clump reabsorbed by the ICM. It is worth noting that such an  equilibrium solution
is allowed only if the cooling of the ICM is accounted for, which was not
in Paper~I. 

The linearised system~\eqref{e-evo}  reads
\bs
\label{e-linear}
\begin{align}
& \der{r}{t} = u
\\
& \der{u}{t} = - g \,\delta
\\
& \der{\delta}{t} = 
\tfrac{3}{5} \; u \; \der{\,\ln K}{r} +  \frac{\omega^\prime}{\tau_{\rm cool}}\; \delta,
\end{align}
\es
where 
\be
\label{e-wprime}
\omega^\prime= \left.\frac{\partial \omega}{\partial \delta} \right |_{\delta=0}
= 2 -  \frac{d\,\ln\Lambda}{d\,\ln T}
\ee
and
\be
\tau_{\rm cool}^{-1}=\tfrac{2}{5}\; \frac{\mu_H\, n_{e}\, \Lambda(T)}{k_B T} 
\ee
is the inverse of the ICM cooling time.

The linearised system admits solutions 
$\delta \propto u \propto {\rm e}^{\lambda\, t}$
if $\lambda$ is the root of the characteristic equation
\bs
\begin{align}
\label{e-char}
& \lambda\left( \lambda^2 - \frac{\omega^\prime}{\tau_{\rm cool}} \lambda +\omega_{\rm BV}^2 \right) =0,
\intertext{where the (squared) Brunt--V\"ais\"al\"a frequency}
&\omega_{\rm BV}^2 = -\tfrac{3}{5}\: {\mathbf g} \cdot \nabla \ln K
=  \tfrac{3}{5}\: \frac{g}{r}  \der{\,\ln K}{\,\ln r}
\end{align}
\es
is positive in a  convectively stable cluster.
For typical parameters $\omega_{\rm BV}\gg\tau_{\rm cool}^{-1}$,
so the eigenvalues are
\be
\lambda=0 
\qquad
\lambda \simeq \omega^\prime/2\,\tau_{\rm cool}  \pm \; i\, \omega_{\rm BV}. 
\ee
One of the eigenvalues is zero, so the equilibrium is non--hyperbolic
and a simple linear analysis does not suffice to characterise the global stability 
of the system (see e.g. \citealp{Gle94}). 
The perturbations oscillate around the equilibrium position 
with the Brunt--V\"ais\"al\"a frequency.
If $\omega'$ (defined by Equation~\ref{e-wprime}) is positive,
the perturbations' amplitude grows  on a cooling time.
The reason of this (overstable) growth is easily understood: 
after the clump has completed an oscillation, its density has grown a little
on account of the  radiative losses, which causes  the amplitude of 
its next oscillation to be larger.   The overstability criterion $\omega'>0$
was already found and discussed by \citet{Bal89}: in particular  
see their equation~(2.1), with $\mathscr L = n^{2}\Lambda(T$).
The oscillations' amplitude  of 
an overstable clump will soon grow so large that it is 
not possible to disregard the non--linear terms of the equation, invalidating
this simple linear analysis.

The global behaviour of the system may be studied in the ``reduced'' set of
equations made by the velocity and overdensity equations only.  In this system
$r$ is  merely a constant parameter.   With a suitable scaling of the variables,
the reduced system reads
\bs
\label{e-reduced}
\begin{align}
& \frac{1}{1+\delta} \; \der{\delta}{t} = u + \xi \; \omega(\delta,T)
\\
& (1+\delta) \; \der{u}{t} = - \eta \; \left(\frac{1+\delta}{1+\delta_{0}}\right)^{1/3} |u| u -\delta,
\end{align}
where
\be
\eta = \tfrac{3}{8}\; C_D \;\frac{g}{a_0\, \omega_{\rm BV}^2}
\qquad
\xi = (\omega_{\rm BV} \, \tau_{\rm cool})^{-1}.
\ee
\es
This system admits the trivial solution $\delta=u=0$:
this is the only proper equilibrium solution of the total system, 
since the velocity must vanish there.   
The non--trivial equilibrium  solutions are
\be
\label{e-redeq}
|u| \, u = - \frac{\delta}{\eta}\left(\frac{1+\delta_{0}}{1+\delta}\right)^{1/3}
\qquad
u  = -\xi \; \omega(\delta, T).
\ee
The first equation  gives  the negative (inwards) velocity
\be
\label{e-terminal}
u_t = - \left[\frac{\delta}{\eta}\left(\frac{1+\delta_{0}}{1+\delta}\right)^{1/3}\right]^{1/2}.
\ee
This is the ``terminal'' velocity of a clump, resulting from the 
equilibrium between the
gravitational pull and the drag force.
The overdensity corresponding to the velocity $u_t$ is implicitly defined 
by  the equation
\be
\label{e-delta2}
\delta_{t}\;\left(\frac{1+\delta_0}{1+\delta_{t}}\right)^{1/3} = \chi \,
\omega_{t}^{2},
\ee
where $\chi = \eta\,\xi^{2}$ and $\omega_{t}\equiv \omega(\delta_{t}, T)$.
This formula will be useful in a moment to derive a simple stability
criterion for the thermal stability of a clump.

\begin{figure}
\begin{center}
\vskip10mm
\includegraphics[width =140mm, angle = 0]{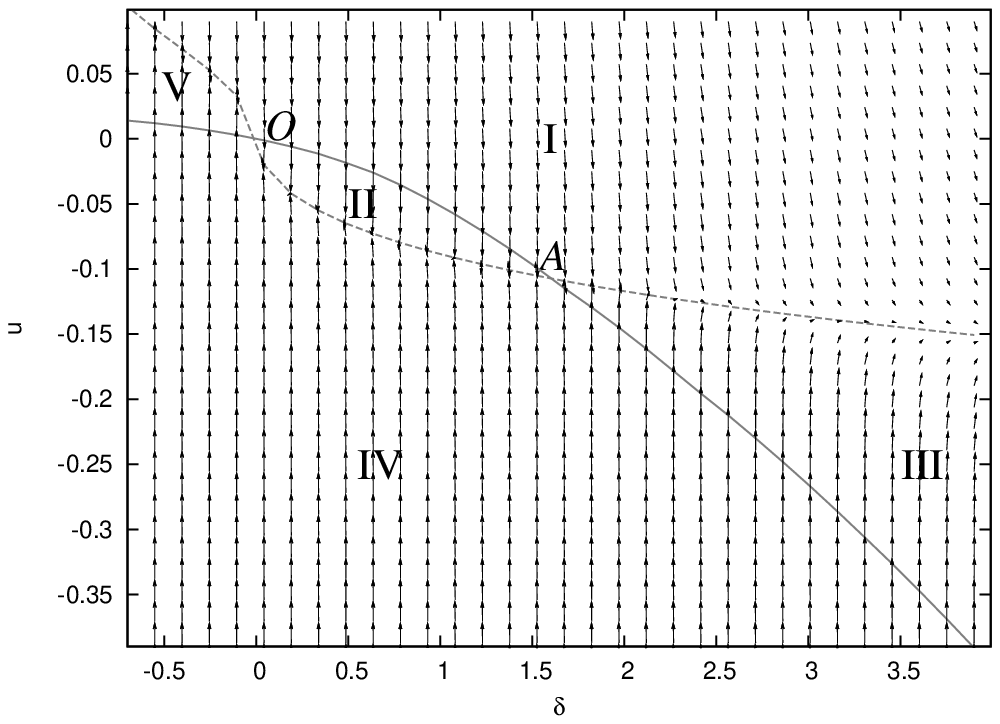} 
\end{center}
\caption{
Vector fields defined by the reduced equation system~\eqref{e-reduced}.
The dimensionless parameters (defined in Appendix~\ref{s-equilibrium}) are 
$\eta = 1.27\times 10^{2}$, 
$\xi  = 1.64\times 10^{-2}$ and
$\chi = 3.42\times 10^{-2}$. They  
correspond to a clump of radius $a_{0}=100\psec$ released from rest at the radius
$r_{0}=20\kpc$ in a cluster like Virgo. 
The points for which  
$d\delta/dt=0$ lie on the solid line, the points for which
$d u/dt=0$ lie on the dashed line.
}
\label{f-dir}
\end{figure}


The  study of the  eigenvalues of the jacobian matrix to assess the stability
of the equilibrium points  $\mathcal O = (0,0)$ and $\mathcal A = (\delta_t, u_t)$ 
is not particularly illuminating.
It is more convenient   to plot (Fig.~\ref{f-dir}) the directions of 
the vector fields $d\delta/dt$ and $d u/d t$ given by~\eqref{e-reduced}.
The lines defined by the equalities $d\delta/dt=0$ (dashed line) and $d u /d t = 0$
(solid line) divide the plane ($\delta-u$) into five regions, 
labelled I, II, III, IV, V.
A point starting from  the upper right of region~I
drifts to the lower right of the plane, 
i.e. its overdensity increases without limits, and its velocity
approaches the terminal velocity~\eqref{e-terminal}.
If the initial position is on the left of region~I,  the trajectory 
crosses vertically the solid line, moves leftwards to region~II, and crosses horizontally the dashed line,  moving to region~IV. 
The trajectory coasts the dashed line, moving to region~V 
and finally to region~I again. The overdensity is 
now smaller than the initial one, and this trajectory winds up around the 
point $\mathcal O$, which is stable.  
This behaviour shows the existence of a 
critical initial overdensity $\delta_{C}$. If $\delta_{0}>\delta_{C}$ 
the clump is
(absolutely) unstable: its overdensity grows more and more, and it is 
eventually accreted on the  cluster centre.

\begin{figure}
\begin{center}
\vskip10mm
\includegraphics[width =110mm, angle = 0]{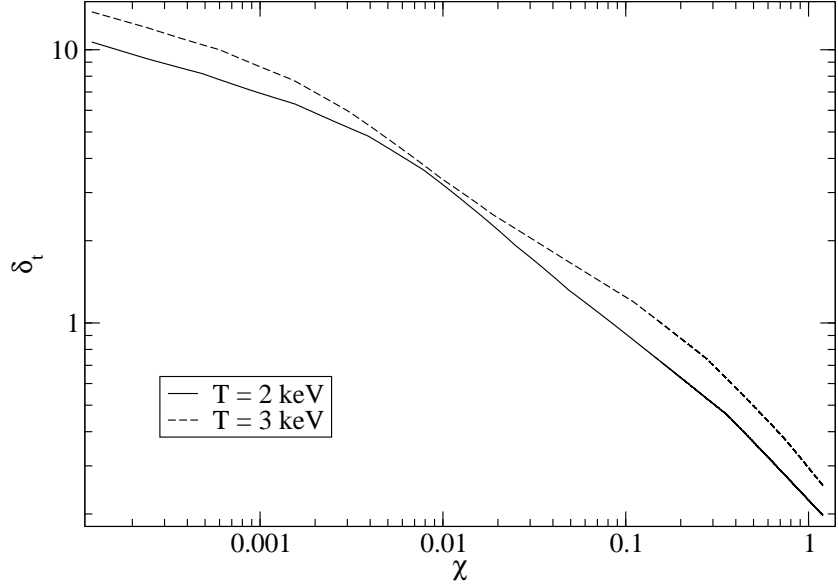} 
\end{center}
\caption{
Critical overdensity $\delta_{t}$, above which constant mass 
clumps are thermally unstable.
$\delta_{t}$ is plotted against the dimensionless stability parameter $\chi$
defined by
Equation~\eqref{e-delta2}. The function $\omega$ 
(defined by~Equation~\ref{e-ft}) has been evaluated at
$T=2.0\keV$ (solid line) and $T=3.0\keV$ (dashed line).
}
\label{f-dcrit}
\end{figure}

We now derive simple formula for the critical density $\delta_{C}$
of  a clump released from rest. 
Consider the trajectory of a clump in the ($\delta-u$) plane.
The starting point of a clump released with zero velocity
is  $\mathcal S=(\delta_0, 0)$.
In a neighbourhood of  $\mathcal S$ the trajectory may be approximated by 
the  straight line
\be
u = u' \, (\delta- \delta_0),
\ee
where
\be
u' = - \frac{\delta_0}{(1+\delta_0)^{2}\,\xi\,\omega_{0}}
\ee
is the derivative $d u/d\delta$ evaluated at $\mathcal S$, and 
$\omega_{0}\equiv\omega(\delta_{0}, T)$.
The line joining $\mathcal S$ and $\mathcal A$ has (negative) angular coefficient
\be
m = \frac{u_{t}}{\delta_{t} - \delta_{0}},
\ee
where $u_{t} = -\xi \,\omega_{t}$, according to the second equation~\ref{e-redeq}.
If $u'=m$ the trajectory originating in  $\mathcal S$ hits the 
point $\mathcal A$: the value of $\delta_{0}$ for which this occurs
corresponds to  the critical overdensity $\delta_{C}$;
if $|u'|<|m|$ the trajectory   
entirely lies in the  region~I of the  ($\delta-u$) plane (Fig.~\ref{f-dir}), and 
if $|u'|<|m|$  the clump moves to region~II, and its  overdensity decreases.
The equation $u'=m$ reads
\be
\label{e-delta3}
\frac{\delta_{C}}{(1+\delta_{C})^{2} \,\xi\,\omega_{C}} 
=
\xi \; \frac{\omega_{t}}{\delta_{t} - \delta_{C}}.
\ee
Equations~\eqref{e-delta2}  (with $\delta_{0} =\delta_{C}$) 
and~\eqref{e-delta3} 
provide a closed system for $\delta_{C}$ and $\delta_{t}$.
This system is quite complex, but it may be simplified
under the assumption $\xi\ll1$, which is true for typical 
parameters.
We may therefore expand
\be\label{e-exp}
\delta_{C} = \delta_{t} + \xi \; \delta_{C}^{(1)} + 
\xi^{2}\; \delta_{C}^{(2)} + o(\xi)^{2}
\ee
To zero order in $\xi$ Equation~\eqref{e-delta3} simply gives
$\delta_{C}=\delta_{t}$, and Equation~\eqref{e-delta2} 
becomes $\delta_{t} = \chi \,\omega_{t}^{2}$.
This equation in $\delta_{t}$ may be solved numerically, with the temperature
$T$ entering as  a parameter.
Plugging the expansion \eqref{e-exp} into Equation~\eqref{e-delta3}
and comparing the coefficients of the powers of $\xi$
we find 
$\delta_{C}^{(1)}=0$
and
$\delta_{C}^{(2)}=\omega_{t}^{2}(1+\delta_{t})^{2}/\delta_{t}$,
to give
\be
\label{e-deltac}
\delta_{C} = \delta_{t} + \xi^{2} \; 
\frac{\omega_{t}^{2} \; (1+\delta_{t})^{2}}{\delta_{t}} + o(\xi)^{2}.
\ee
Since $\xi\simeq 10^{-2}\ll1$, the zero--order approximation
is usually  satisfactory. In this case
the value $\delta_{C}\simeq\delta_{t}$ only depends on the dimensionless
parameter $\chi$; a couple of examples are
shown in  Fig.~\ref{f-dcrit} for different temperatures. 

\begin{figure}
\begin{center}
\vskip10mm
\includegraphics[width =110mm, angle = 0]{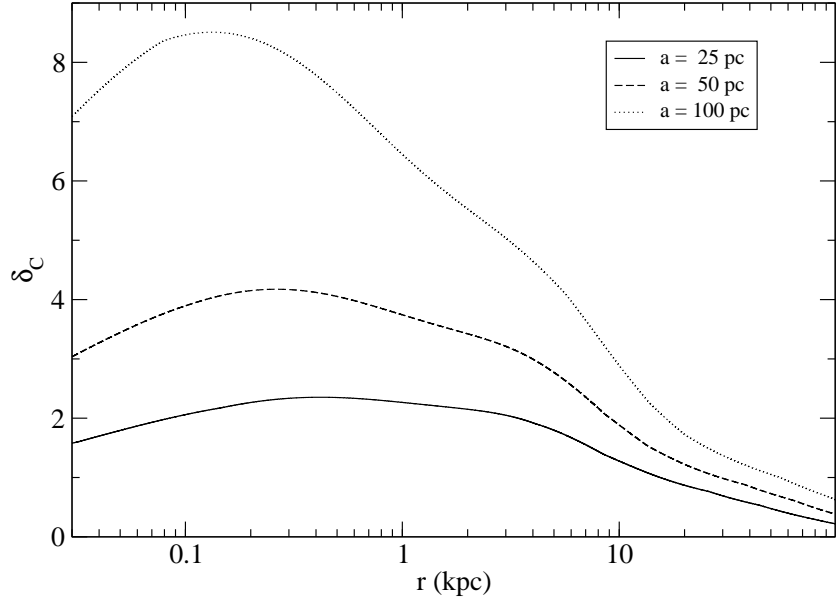} 
\end{center}
\caption{
Critical overdensity $\delta_{C}$, above which constant mass 
clumps are thermally unstable as a function of the distance from the 
centre, for Virgo.
The curves refer to different initial radii of the clumps:
$a_{0}=25\psec$ (solid line);
$a_{0}=50\psec$ (dashed line) and
$a_{0}=100\psec$ (dotted line).
Clumps with $\delta>\delta_{C}$ are accreted, the other are thermally
stabilised at $r>0$.
}
\label{f-dtr}
\end{figure}
Fig.~\ref{f-dtr}  shows  the radial profile of $\delta_{C}$ ---as
calculated from the second--order formula~\eqref{e-deltac}---
for some values of the clumps' initial radius $a_{0}$.
The radial  temperature, density and entropy profiles  
are those of the Virgo cluster.

A final warning  about this stability analysis is in order. 
In the present analysis we have neglected the motion of the clump through the 
cluster:
the radius entered only implicitly into the coefficients $\eta$ and $\xi$,
treated as constants. If the radial motion of the  clump is considered, 
the situation is slightly more involved.
First, the clump slightly drifts inwards, 
and its evolution occurs at a smaller distance to the centre, 
where $\delta_{C}$ is larger (if $r\ga 0.2\kpc$, Fig.~\ref{f-dtr}).
The value of $\delta_{C}$ found with the analysis of the reduced 
system~\eqref{e-reduced} is therefore slightly underestimated. 
Second,
even a clump with $\delta<\delta_{C}$ may be accreted to the centre:
we term this behaviour as ``relative instability''. 
If $\delta_{0}$ is not much  smaller than
$\delta_{C}$, the clump's overdensity decreases, but the clump may
nevertheless  reach the centre  before the equilibrium value 
$\delta=0$ is attained.
The value $\delta_{C}$ calculated here  may only distinguish  between
the ``relative instability'' (for which $\dot\delta<0$, with 
{\em possible} eventual accretion) and  the ``absolute instability'',
for which $\dot\delta>0$ all the way.


\clearpage

\bsp

\label{lastpage}

\end{document}